\documentclass[a4paper,aps,prd,twocolumn,noshowpacs,preprintnumbers,nofootinbib]{revtex4-1}

\usepackage[utf8x]{inputenc}
\usepackage{epsfig,amssymb,amsmath,psfrag}

\usepackage{etex}
\usepackage{pstricks-add}
\usepackage{verbatim}

\usepackage{bm}
\usepackage{epic}
\usepackage{graphicx}
\usepackage{color}
\usepackage{hyperref}
\usepackage{young}
\usepackage{multirow}
\usepackage{fancybox}

\usepackage[paperwidth=210mm,paperheight=297mm,centering,hmargin=1.75cm,vmargin=2.5cm]{geometry}

\def \be  {\begin{equation}}
\def \ee  {\end{equation}}
\def \ba  {\begin{eqnarray}}
\def \ea  {\end{eqnarray}}
\def \baa {\begin{eqnarray*}}
\def \eaa {\end{eqnarray*}}
\def \bb  {\begin {thebibliography} }
\def \eb  {\end{thebibliography}}
\def \lab #1 {\label{#1}}

\newcommand {\non}{\nonumber}

\def\Tr{\mathrm{Tr}}

\def\bseq{\begin{subequation}}  
\def\eseq{\end{subequation}}
\def\bsea{\begin{subeqnarray}}  
\def\esea{\end{subeqnarray}}

\newcommand{\beq}{\begin{equation}}
\newcommand{\bea}{\begin{eqnarray}}
\newcommand{\eea}{\end{eqnarray}}
\newcommand{\eeq}{\end{equation}}

\renewcommand{\a}{\alpha}

\newcommand{\g}{\gamma}
\newcommand{\G}{\Gamma}

\newcommand{\e}{\epsilon}

\newcommand{\s}{\sigma}

\begin{document}

\title{Probing Wilson loops in $\mathcal{N}=4$ Chern--Simons--matter theories at weak coupling}

\author{Luca Griguolo${}^{1}$}
\author{Matias Leoni${}^{2}$}
\author{Andrea Mauri${}^{3}$}
\author{Silvia Penati$^{3,4}$} 
\author{Domenico Seminara${}^{5}$}

\affiliation{\vskip 7pt  ${}^1${Dipartimento di Fisica e Scienze della Terra, Universit\`a di Parma and INFN Gruppo Collegato di Parma, Viale G.P. Usberti 7/A, 43100 Parma, Italy}\\
${}^2${Physics Department, FCEyN-UBA \& IFIBA-CONICET\ Ciudad Universitaria, Pabell\'on I, 1428, Buenos Aires, Argentina }\\
${}^3${Dipartimento di Fisica dell'Universit\`a degli studi di Milano--Bicocca, Piazza della Scienza 3, I-20126 Milano, Italy}\\
${}^4${INFN Sezione di Milano--Bicocca, piazza della Scienza 3, I-20126 Milano, Italy}\\
${}^5${Dipartimento di Fisica, Universit\`a di Firenze and INFN Sezione di Firenze, via G. Sansone 1, 50019 Sesto Fiorentino, Italy}\\
{\tt  luca.griguolo@pr.infn.it, leoni@df.uba.ar, andrea.mauri1@unimib.it, silvia.penati@mib.infn.it, domenico.seminara@fi.infn.it}}

\begin{abstract}
For three--dimensional ${\cal N}=4$ super Chern--Simons--matter theories associated to necklace quivers $U(N_0) \times U(N_1) \times \cdots  U(N_{2r-1})  $, we study at quantum level the two kinds of 1/2 BPS Wilson loop operators recently introduced in \href{http://arxiv.org/abs/1507.}{arXiv:1506.07614}. We perform a two--loop evaluation and find the same result for the two kinds of operators, so moving to higher loops a possible quantum uplift of the classical degeneracy. We also compute the 1/4 BPS bosonic Wilson loop and discuss the quantum version of the cohomological equivalence between fermionic and bosonic Wilson loops. We compare the perturbative result with the Matrix Model prediction and find perfect matching, after identification and remotion of a suitable framing factor. Finally, we discuss the potential appearance of three--loop contributions that might break the classical degeneracy and briefly analyse possible implications on the BPS nature of these operators. 
\\[0.3cm]
PACS numbers: 11.25.Hf, 11.30.Pb, 11.15.Yc, 11.10.Kk \vspace{-0.3cm}
\end{abstract}


\keywords{BPS Wilson loops, Chern--Simons--matter theories, localization}

\maketitle 

\section{Introduction}
 
One of the most interesting classes of observables in supersymmetric gauge theories is constitued by BPS Wilson loops \cite{Erickson:2000af,Drukker:2000rr}. They provide an exciting arena where exact computations can be performed through localization techniques \cite{Pestun:2007rz}, so interpolating non-trivially between weak and strong coupling regimes. 

The first and most famous example is the 1/2 BPS circular Wilson loop, originally constructed in ${\cal N}=4$ super Yang-Mills theory. It is calculated by a simple Gaussian matrix model and reproduced at strong coupling through the AdS/CFT correspondence \cite{Erickson:2000af,Drukker:2000rr}. The original proposal has been generalized to less supersymmetric loops \cite{Drukker:2007qr} and in theories with  ${\cal N}=2$ supersymmetry \cite{Pestun:2007rz}. In all these constructions the key point is to improve the holonomy of the gauge connection by coupling some of the scalar fields to the appropriate contours. The resulting operators are BPS and their expectation values can be computed by adding a suitable Q-exact term to the classical action, so that the relevant path-integral is semiclassically exact \cite{Pestun:2007rz}. 

In three dimensions the story is a little bit different. ${\cal N}=2$ Chern-Simons theories still possess circular 1/2 BPS Wilson loops obtained through scalar couplings, which are calculated by localization techniques \cite{Kapustin:2009kz}. Going to more supersymmetric theories, as the ${\cal N}=6$ ABJ(M) model, the construction of 1/2 BPS operators has to be refined \cite{Drukker:2009hy} (see also \cite{Cardinali} for a generalization to other contours). In fact, scalar couplings only provide 1/6 BPS Wilson loops and fermionic couplings have to be invoked to enhance supersymmetry. More surprisingly, 1/2 BPS Wilson loops in ABJ(M) theory are seen equivalent to a linear combination of 1/6 BPS ones \cite{Drukker:2009hy,Drukker:2010nc}. In fact, they belong to the same cohomology class of the localizing supercharge and thus, up to framing anomalies, they are the same observable at quantum level. This phenomenon is a three-dimensional novelty, that has been checked concretely in perturbation theory \cite{Bianchi:2013rma,Griguolo:2013sma} and certainly needs a more profound investigation. Recently, the construction of 1/2 BPS Wilson loops has been presented \cite{Ouyang:2015qma, Cooke:2015ila} in ${\cal N}=4$ quiver Chern-Simons theories \cite{Gaiotto:2008sd, Hosomichi:2008jd}.
In the case of circular and linear quivers with non-vanishing CS levels, two apparently independent 1/2 BPS circular loops emerge, which share the same supersymmetry and belong to the same cohomology class of the familiar bosonic 1/4 Wilson loop operator. These properties have been derived at classical level and should be checked against truly quantum computations, where divergences and/or anomalies could arise, possibly lifting the classical degeneracy.

 In this paper we perform explicitly a perturbative computation of the two fermionic Wilson loops at second order in the coupling constant, finding perfect consistency with the classical picture and no lifting of the quantum expectation value. At the same order we check  the matrix model result obtained from the localization procedure and, consequently, confirm the cohomological equivalence with the 1/4 BPS loop. The plan of our Letter is the following. In Section \ref{section2} we briefly recall the construction of the Wilson loop operators in ${\cal N}=4$ circular Chern-Simons quivers. Section \ref{section3} is devoted to the perturbative computation of the expectation value of the relevant Wilson loop operators. In Section \ref{section4} we check their cohomological equivalence at quantum level. Matrix model results are explictly seen to be consistent with our quantum calculations in Section \ref{section5}. A critical analysis of the degeneracy problem is presented in Section \ref{section6}, where we discuss the potential appearance of higher--order contributions that might turn out to be different for the two fermionic Wilson loops.

\section{Circular BPS Wilson loops in ${\cal N}=4$ CS--matter theories}
\label{section2}

We consider a Chern--Simons--matter theory associated to a circular quiver with gauge group $U(N_0) \times U(N_1) \times \cdots U(N_{2r-1})$ ($N_{2r} \equiv N_0$).  Besides the gauge sector containing vectors $A_{(A)}^\mu$ in the adjoint representation of the group $U(N_A)$, the theory contains matter scalars $(q_{(2A+1)}^I)^j_{\; \hat{j}}$ ($(\bar{q}_{(2A+1) I})^{\hat{j}}_{\; j}$) in the (anti)bifundamental representation of the $U(N_{2A+1})$, $U(N_{2A+2})$ nodes (indices $j$ and $\hat j$, respectively) and in the fundamental of the R-symmetry $SU(2)_L$ ($I=1, 2$), twisted scalars $(q_{(2A)}^{\hat I})^{\hat j}_{\; j}$ ($(\bar{q}_{(2A) \hat I})^{j}_{\; \hat{j}}$) in the (anti)bifundamental representation of $U(N_{2A})$, $U(N_{2A+1})$ nodes and in the  fundamental of  
the R-symmetry $SU(2)_R$ ($\hat I=1, 2$), plus the corresponding fermions $(\psi_{(2A+1) \hat I})^j_{\; \hat{j}}$ ($(\bar{\psi}_{(2A+1)}^{\hat I})^{\hat{j}}_{\; j}$) and $(\psi_{(2A) I})^{\hat{j}}_{\; j}$ ($(\bar{\psi}_{(2A)}^I)^{j}_{\; \hat{j}}$), respectively.

In three--dimensional euclidean space the classical action reads 
\beq
S = \sum_{A=0}^{2r-1} \left( S_{CS}^{(A)} + S_{mat}^{(A)} \right)  + S_{pot} + S_{gf}  
\eeq
where
\bea
\label{action}
&& S_{CS}^{(A)} = - \frac{i}{2} k_A \, \int d^3x\,\varepsilon^{\mu\nu\rho}  {\Tr} \Big( A_{(A)\mu} \partial_\nu A_{(A)\rho} 
\\
&~& \qquad \qquad \qquad \qquad \qquad \quad +\frac{2}{3} i A_{(A)\mu} A_{(A)\nu} A_{(A)\rho} \Big)
\non \\
\non \\
&& S_{mat}^{(A)} = \int d^3x \, {\Tr} \Big[  D_\mu q_{(A)} D^\mu \bar{q}_{(A)}+ i \, \bar{\psi}_{(A)}  \g^\mu D_\mu \psi_{(A) } \Big]
\non 
\eea
while $S_{gf}$ is the gauge--fixing plus ghost action and $S_{pot}$ the matter interaction action, whose explicit expression can be found for instance in \cite{Imamura:2008dt}. 
This part of the action does not enter two--loop diagrams, so we will ignore it in the rest of the paper.  

${\cal N}=4$ supersymmetry requires the CS levels to satisfy
\begin{equation}
k_A=\frac{k}{2}(s_A-s_{A-1}),\qquad
s_A=\pm1,\qquad
k>0
\label{ki}
\end{equation}
We will consider the case $s_A=(-1)^{A+1}$, which leads to alternating $\mp k$ levels.

In \cite{Ouyang:2015qma, Cooke:2015ila} Wilson loop operators (WL) have been introduced that are classically BPS. These are defined locally for each site of the quiver and involve at most three adjacent nodes. Therefore, restricting for simplicity to node $U(N_1)$ and its nearest--neighbour $U(N_0)$ and $U(N_2)$ we will consider the following loop operators integrated on the unit circle $\G$ ($x^\mu = (\cos{\tau}, \sin{\tau}, 0)$, $\tau \in [0,2\pi]$):

\vskip 10pt
\noindent
\underline{Fermionic 1/2 BPS $\psi_1$--Wilson loop}. When referred to node $U(N_1)$ it is defined as \cite{Cooke:2015ila}
\beq
W_{\psi_1}[\G] = \frac{1}{N_1 + N_2} \, {\Tr} \,P \exp{ \left( -i \int_\G d\tau {\cal L}^{\psi_1}_F(\tau)\right) } 
\eeq
where  
\bea \label{WL1}
&& {\cal L}^{\psi_1}_F = \left( \begin{array}{cc} {\cal A}_{(1)}  & \bar c_\a \psi_{(1) \hat 1}^\a \\ c^\a \bar \psi^{\hat 1}_{(1) \a}  &  {\cal A}_{(2)} \end{array} \right) 
\\
&& {\cal A}_{(1)} =  \dot x^{\mu} A_{(1) \mu}   - \frac{i}{k} \left( q_{(1)}^I \delta_I^{\; J}  \bar q_{(1)  J} + \bar q_{(0) \hat I}  (\sigma_3)^{\hat I}_{\; \hat J}  \, q_{(0)}^{\hat J} \right)
\non \\
&& {\cal A}_{(2)} =    \dot x^{\mu} A_{(2) \mu}   - \frac{i}{k} \left(  \bar q_{(1) I}  \delta^I_{\; J} \, q_{(1)}^{ J}+  q_{(2)}^{\hat I} (\sigma_3)_{\hat I}^{\; \, \hat J} \bar q_{(2) \, \hat J}  \right)
\non
\eea
and the commuting spinors $c, \bar c$ are (with $C \bar C = -\tfrac{i}{k}$)
\bea \label{c1}
&& c(\tau) = C ( \cos{\tfrac{\tau}{2}} - \sin{\tfrac{\tau}{2}}, \cos{\tfrac{\tau}{2}} + \sin{\tfrac{\tau}{2}})
\non \\
&& \bar{c}(\tau) =  \bar C \left( \begin{array}{c}  
\cos{\frac{\tau}{2}} - \sin{\frac{\tau}{2}}  \\
\cos{\frac{\tau}{2}} + \sin{\frac{\tau}{2}}
 \end{array}\right)  
 \eea
 
\vskip 10pt
\noindent
\underline{Fermionic 1/2 BPS $\psi_2$--Wilson loop}. This loop operator is defined as \cite{Cooke:2015ila}
\beq
W_{\psi_2} [\G] = \frac{1}{N_1 + N_2}  \, \Tr \,P \exp{ \left( -i \int_\G d\tau {\cal L}^{\psi_2}_F(\tau)\right) } 
\eeq
where 
\bea\label{WL2}
&& {\cal L}^{\psi_2}_F = \left( \begin{array}{cc} {\cal A}_{(1)}  & \bar c_\a \psi_{(1) \hat 2}^\a \\ c^\a \bar \psi^{\hat 2}_{(1) \a}  &  {\cal A}_{(2)} \end{array} \right) 
\\
&& {\cal A}_{(1)} =   \dot x^{\mu} A_{(1) \mu}   - \frac{i}{k} \left( -q_{(1)}^I \delta_I^{\; J}  \bar q_{(1)  J} + \bar q_{(0) \hat I}  (\sigma_3)^{\hat I}_{\; \hat J}  \, q_{(0)}^{\hat J} \right)
\non \\
&& {\cal A}_{(2)} =   \dot x^{\mu} A_{(2) \mu}   - \frac{i}{k} \left( - \bar q_{(1) I}  \delta^I_{\; J} \, q_{(1)}^{ J}+  q_{(2)}^{\hat I} (\sigma_3)_{\hat I}^{\; \, \hat J} \bar q_{(2) \, \hat J}  \right)
\non
\eea
and the commuting spinors $c, \bar c$ given by (with $C \bar C = \tfrac{i}{k}$)
\bea
\label{c2}
&& c(\tau) =   - C (\cos{\tfrac{\tau}{2}} + \sin{\tfrac{\tau}{2}}, -\cos{\tfrac{\tau}{2}} + \sin{\tfrac{\tau}{2}}) 
\non \\
&& \bar{c}(\tau) =  \bar C \left( \begin{array}{c}  
 \cos{\frac{\tau}{2}} + \sin{\frac{\tau}{2}} \\
-\cos{\frac{\tau}{2}} + \sin{\frac{\tau}{2}}
 \end{array}\right)  
 \eea
This loop differs from the previous one for the replacement of the identity matrix with minus the identity matrix in the scalar couplings,  the replacement $\psi_{(1)}^{\hat 1} \to \psi_{(1)}^{\hat 2}$ in the off--diagonal elements and the choice of different fermion couplings. 

\vskip 10pt
\noindent
\underline{Bosonic 1/4 BPS Wilson loop}. We will  be also interested in bosonic loop operators that respect 1/4 of the original supersymmetries \cite{Ouyang:2015qma, Cooke:2015ila}. For sites $N_1$ and $N_2$  they are    
\bea
\label{bosonicWL}
&& W^{(1)}[\G] = \frac{1}{N_1} \, {\Tr} \,P \exp{ \left( -i \int_\G d\tau {\cal L}_B^{(1)}(\tau)\right) } 
\non \\
&& W^{(2)}[\G] = \frac{1}{N_2} \, {\Tr} \,P \exp{ \left( -i \int_\G d\tau {\cal L}_B^{(2)}(\tau)\right) } 
\eea
where
\bea
&& {\cal L}_B^{(1)} =\dot x^\mu A_{(1) \mu} - \frac{i}{k} \left(\bar q_{(0) \hat I} (\sigma_3)^{\hat I}_{\; \hat J}  q_{(0)}^{\hat J}  +
q_{(1)}^I (\sigma_3)_I^{\;\;J} \bar q_{(1) \, J}  \right)  
\non \\ 
&& {\cal L}_B^{(2)} = \dot x^\mu A_{(2) \mu} - \frac{i}{k} \left(\bar q_{(1) I} (\sigma_3)^I_{\; J}  q_{(1)}^{J}  +
q_{(2)}^{\hat I} (\sigma_3)_{\hat I}^{\; \hat J}  \bar q_{(2) \, \hat J}  \right)   
\non
\eea

\vskip 10pt
As proved in \cite{Ouyang:2015qma, Cooke:2015ila} the fermionic Wilson loops are classically equivalent to the bosonic ones,  
\beq
\label{coho}
W_{\psi_i} = \frac{N_1 W^{(1)} + N_2 W^{(2)}}{N_1 + N_2} + Q V_{\psi_i}  \qquad i=1,2
\eeq
up to a $Q$-term, where $Q$ is some linear combination of supercharges. If this cohomogical equivalence survives at quantum level,  localization techniques applied to the bosonic Wilson loops provide an all--order prediction also for the fermionic operators. For the ABJM orbifold case ($N_i \equiv N$ for any $i$) the corresponding matrix model has been computed in \cite{Ouyang:2015hta}.

\section{Two--loop evaluation} \label{section3}

In this Section, we present the results for the circular $1/2$ BPS and $1/4$ BPS WL up to two loops. The computation, that requires regularizing UV divergences and evaluating intricate trigonometric integrals, heavily relies on the techniques introduced in \cite{Bianchi:2013rma, Griguolo:2013sma} to which we refer for details. 

We use dimensional regularization with dimensional reduction (DRED) to control potentially divergent integrals. They generally converge in the complex half--plane defined by some critical value of the real part of the regularization parameter $\epsilon$. Using techniques described in \cite{Bianchi:2013rma}, they can be computed analytically for any complex value of $\epsilon$ and turn out to be expressible in terms of hypergeometric functions. Their actual value for $\epsilon \to 0$ can be then obtained by analytically continuing the hypergeometric functions close to the origin and expanding the result up to finite terms.

At one--loop we have only two contributions associated with the exchange of one gluon and one fermion line, respectively. The vector exchange vanishes because of the planarity of the circular contour that gets contracted with the Levi--Civita tensor. The contribution from the fermion exchange is proportional to
\begin{align} \label{1Loop}
\int_0^{2\pi} d\tau_1 \int_0^{\tau_1} d\tau_2 \; \frac{(c_1 \g^\mu \bar{c}_2) \, (x_1 - x_2)_\mu}{[(x_1 - x_2)^2]^{\frac32-\e}} 
\end{align}
Choosing the set of euclidean gamma matrices $\g^\mu = \{\s^3, \s^1,\s^2 \}$, and taking into account the explicit expression of fermion couplings (\ref{c1}) we can write 
\beq
\label{vector}
(c_i \g^\mu \bar{c}_j)  (x_i - x_j)_\mu = - \tfrac{4i}{k} \sin{\tfrac{\tau_1 - \tau_2}{2}}
\eeq
Therefore, the integral becomes  
\beq
\label{IntegralW12}
\int d\tau_{1>2}   \; \frac{1}{[\sin^2{\frac{\tau_{12}}{2}}]^{1-\e}}  = \frac{2 \pi ^{3/2} \Gamma \left(-\frac{1}{2}+\epsilon \right)}{\Gamma \left(\epsilon \right)}
\eeq
and this expression vanishes in 3d. Therefore, we do not have any one--loop contribution to $W_{\psi_1}$. 

\begin{figure}
\centering
 \includegraphics[width=0.40\textwidth]{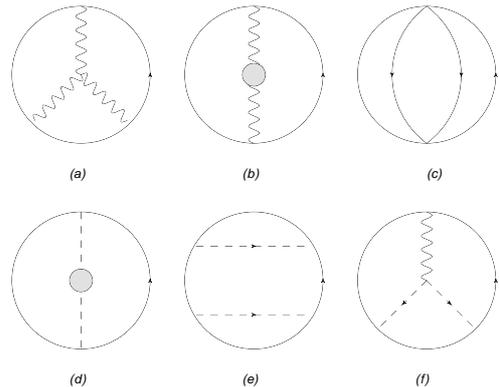}
\caption{Non-vanishing two--loop diagrams for  $\psi_1$ and $\psi_2$ loops. Wavy lines represent gauge propagators, solid lines represent scalars, and dashed lines are fermion propagators. Bubbles represent one--loop corrections to the propagators, as given in Appendix. }
    \label{2loop}
\end{figure}

We then move to two loops. Contributions that are not trivially vanishing for planarity of the contour are associated to the diagrams in Fig. \ref{2loop}. 
We list the results for each single diagram, while for details we refer the reader to \cite{Bianchi:2013rma, Griguolo:2013sma}. 

\vskip 10pt
\noindent
\underline{Diagram (a)} - It comes from the gauge part of the third order expansion of the WL contracted with the gauge cubic vertex. Summing the contributions from the two connections $A_{(1)}$ and $A_{(2)}$, we have  
\beq
\label{int}
{\rm (a)}_{\psi_1} =  - \frac{ N_1(N_1^2-1)+N_2(N_2^2-1) }{N_1+N_2} \, \frac{1}{k^2} \, 
\frac{\G^3(\frac32 - \e)}{8\pi^{\frac{9}{2} - 3\e}} \, I^{(a)}
\eeq
where  
\bea
&& I^{(a)} =  \int d\tau_{1>2>3} \, \dot{x}_1^\s \dot{x}_2^\eta \, \dot{x}_3^\zeta \, \varepsilon^{\xi \tau \kappa} \varepsilon_{\s\xi\mu} \varepsilon_{\eta\tau\nu} \varepsilon_{\zeta \kappa \rho}  \times
\non \\ 
&&\int d^{3-2\e} x    
\frac{  (x-x_1)^\mu (x-x_2)^\nu (x-x_3)^\rho}{|x-x_1|^{3-2\e} |x-x_2|^{3-2\e} |x-x_3|^{3-2\e} } 
\eea
This integral, being finite, can be computed at $\e =0$ and eventually gives $I^{(a)} = \frac{8}{3} \pi^3$ \cite{Rey:2008bh, Bianchi:2013rma}. The final result is then
\beq
\label{a1}
{\rm (a)}_{\psi_1} =   - \frac{1}{24} \, \frac{N_1^2 + N_2^2 - N_1N_2 - 1}{k^2}   
\eeq

\vskip 10pt
\noindent
\underline{Diagrams (b) + (c)} - Summing the two contributions we obtain
\bea
\label{bc}
&& [{\rm (b) + (c)}]_{\psi_1} =
\\
&~& \frac{N_1^2(N_0+N_2) + N_2^2(N_1+N_3)}{N_1+N_2} \, \frac{1}{k^2} \, \frac{\G^2(\frac12-\e)}{8\pi^{3 -2\e}}  \,  I^{(b+c)}
\non
\eea
where \cite{Rey:2008bh}
\beq
I^{(b+c)} = \int d\tau_{1>2} \, \frac{- \dot{x}_1 \cdot \dot{x}_2 + |\dot{x}_1| |\dot{x}_2|}{[(x_1-x_2)^2]^{1-2\e}} \underset{\e \rightarrow 0}{\longrightarrow}
  \pi^2
\eeq
We then obtain
\beq
[{\rm (b) + (c)}]_{\psi_1} =   \frac{N_1^2(N_0+N_2) + N_2^2(N_1+N_3)}{N_1+N_2} \, \frac{1}{8k^2} 
\eeq

\vskip 10pt
\noindent
\underline{Diagram (d)} -  This contribution is proportional to the exchange of a one--loop fermion propagator. Using its explicit expression given in \cite{Bianchi:2013rma} we obtain
\beq
\label{d1}
{\rm (d)}_{\psi_1} \sim   \int d\tau_{1>2} \, \frac{|\dot{x}_1| |\dot{x}_2|}{[(x_1-x_2)^2]^{1-2\e}} \, \left[ (c_1 \bar{c}_2) - (c_2 \bar{c}_1) \right] 
\eeq
where we indicate $c_i \equiv c(\tau_i)$ on the circle. As follows from eq. (\ref{c1}) we have 
\beq
\label{scalar}
(c_i \bar{c}_j) = -\tfrac{2i}{k} \cos{\tfrac{\tau_i - \tau_j}{2}}
\eeq
so that diagram (d) vanishes identically.

\vskip 10pt
\noindent
\underline{Diagram (e)} - Expanding the $\psi_1$--loop at forth order and performing the two possible contractions of fermions we obtain a linear combination of terms of the form
\beq
\label{e2}
 \frac{\G^2(\frac32 -\e)}{4 \pi^{3-2\e}}   \!\! \int \!\! d\tau_{1>..>4} \frac{(c_i \g^\mu \bar{c}_j) (c_k \g^\nu \bar{c}_l)  (x_i - x_j)_\mu (x_k - x_l)_\nu }{[(x_i - x_j)^2 (x_k - x_l)^2]^{\frac32 - \e}}
\eeq
 This expression can be easily elaborated by using identity (\ref{vector}) twice. The contour integrals we are left with are divergent. They can be evaluated away from $\e = 0$ and suitably continued close to the origin (see \cite{Bianchi:2013rma} for details). The result is 
\beq
{\rm (e)}_{\psi_1} = \frac38 \, \frac{N_1 N_2}{k^2}
\eeq
   
\vskip 10pt
\noindent
\underline{Diagram (f)} - Expanding $W_{\psi_1}$ at third order and contracting with one mixed vertex $\bar \psi \psi A$ coming from the action, we obtain the linear combination of six integrals of the form 
\beq
\label{first}
\frac{1}{k}  \, \frac{\G^3(\frac12 -\e)}{64 \pi^{\frac92 - 3\e}} \int d\tau_{1>2>3}  
( \bar{c}_i \g^\xi \g^\mu \g^\s c_j) \, \dot{x}_k^\nu \, \varepsilon_{\nu \mu}^{\phantom{\nu \mu} \rho} \, \G_{\rho \xi \sigma}
 \eeq
where the labels $i,j,k$ run over $1,2,3$ and
\bea
\label{gammaintegrals}
\Gamma^{\mu\nu\rho}  \, = \,
\partial^{\mu}_k\, \partial^{\nu}_i\, \partial^{\rho}_j\, \int \frac{d^3 x}{[ (x-x_1)^2 (x-x_2)^2 (x-x_3)^2]^{\frac12 -\e}} 
\non
\eea
The spinorial structure appearing in (\ref{first}) can be simplified by using standard identities for the product of three Pauli matrices. It is easy to prove that, because of the planarity of the contour the only non--vanishing contributions we are left with are proportional to the bilinears $(c_i \bar c_j)$ and $(c_i \gamma^3 \bar c_j)$.  Using identity (\ref{scalar}) together with
\beq
(c_i \g^3 \bar{c}_j)   = \tfrac{2}{k} \sin{\tfrac{\tau_1 - \tau_2}{2}} \quad ,
\eeq
computing the corresponding color factors and evaluating the integrals using the procedure described in \cite{Bianchi:2013rma} we finally obtain
\beq
{\rm (f)}_{\psi_1} = - \frac12 \, \frac{N_1N_2}{k^2}   
\eeq

Summing all the contributions the two--loop result for the $\psi_1$--loop is 
\bea
 \label{2loopresult}
&& \langle W_{\psi_1}[\G]  \rangle|_{\rm 2loop}    =   
\\
&&1 -\frac{1}{24 k^2} \left[ (N_1^2 + N_2^2 - N_1 N_2 -1) - 3 \frac{N_0 N_1^2 + N_3 N_2^2}{N_1+N_2} \right]
\non
\eea

We now consider the Wilson loop  $W_{\psi_2}$ defined in (\ref{WL2}, \ref{c2}). Its perturbative evaluation can be easily performed by exploiting the previous results, where we should take into account that the $\psi_2$--loop has slightly different scalar couplings in the ${\cal A}$--terms and different fermionic couplings $c, \bar c$. The fact that the $\psi_{(1) \hat 2}$ fermion replaces $\psi_{(1) \hat 1}$ does not make much difference, as the tree--level propagator for the two fermionic components is the same.

At one loop, the fermion exchange diagram (see equation (\ref{1Loop})) involves the bilinear $(c_i \g^\mu \bar{c}_j)  (x_i - x_j)_\mu $. Computing it with the assignment (\ref{c2}), we obtain the same result (\ref{vector}) up to an overall sign.  However, since the diagram is still proportional to integral (\ref{IntegralW12}),  the $\psi_2$--loop contribution at one loop also vanishes in the $\e \to 0$ limit.

At two loops, non--vanishing contributions are still given in Fig. \ref{2loop}. It is easy to argue that the first three bosonic diagrams give the same result as $W_{\psi_1}$. In fact, diagram (a) and (b) involve only gauge fields, so they are insensitive to changes in matter couplings. In diagram (c) the matrices ($I$ and $\s_3$) governing the scalar couplings enter quadratically, so that the sign difference between the two WL definitions does not affect the calculation.
Changes in the calculation might be expected from diagrams containing fermions, since a different set of fermionic couplings may give rise to different expressions for the fermionic bilinears $(c_i \bar c_j)$ and $(c_i \gamma^\mu \bar c_j)$. However, the contribution from diagram (d)  is still proportional to expression (\ref{d1}) and vanishes since, as before, $(c_1 \bar{c}_2) = (c_2 \bar{c}_1)$, as follows immediately from  (\ref{c2}).  In diagram (e) the double fermion contractions read again as in eq. (\ref{e2}), which involves the bilinear $(c_i \g^\mu \bar{c}_j)  (x_i - x_j)_\mu $. As we already mentioned, this bilinear has an overall sign compared to the corresponding expression for $W_{\psi_1}$. However, in (\ref{e2}) the product of two such expressions appears, so that the final result is the same as for $W_{\psi_1}$. Finally, diagram (f) only involves minimal coupling of fermions to the gauge vectors, which is identical for  $\psi_{(1) \hat 1}$ and  $\psi_{(1) \hat 2}$. Therefore the evaluation of the integrals still depends on the spinorial bilinears $(c_i \bar c_j)$ and $(c_i \g^3 \bar c_j)$. Using (\ref{c2}), these can be quickly shown  to be identical to the ones for the $\psi_1$--loop. Here it is crucial that, due to the planarity of the contour, only the bilinear $(c_i \g^{\mu} \bar c_j)$ with $\mu=3$ enters the calculation. If this were not the case, we would obtain a different result, since for the bilinears along the directions $\mu=1,2$ where the circular contour lies there is a sign difference between the two WL.  

Summarizing, we find that  
 \bea
\langle W_{\psi_1 }[\G]  \rangle|_{\rm 2loop}    =  \langle W_{\psi_2}[\G]  \rangle|_{\rm 2loop}  
\eea
Therefore, up to this order, there is no quantum uplift of the degeneracy between the two fermionic WL.

Exploiting the previous calculation, it is also immediate to determine the $1/4$ bosonic WLs (\ref{bosonicWL}). Again, there is no one--loop contribution, while the two--loop ones are given by the first three diagrams in Fig. 1. With suitable adjustments we find ($A=1,2$)
\bea
\label{bosonicresult}
\langle W^{(A)} \rangle = 1 - \frac{1}{24 k^2} \left[ N_A^2 - 3 N_{A-1} N_A - 3 N_A N_{A+1} -1 \right] 
\non \\
\eea
Note that, under identification $N_0=N_2$ and $N_3=N_1$,  our results (\ref{2loopresult}, \ref{bosonicresult}) coincide with the two--loop expressions for the $1/2$ and $1/6$ Wilson loops in ABJ, respectively \cite{Drukker:2009hy, Rey:2008bh, Bianchi:2013rma, Griguolo:2013sma}. Moreover, in the orbifold ABJM $[U(N) \times U(N)]^r$ ($N_{A} = N$ for all the nodes) the result becomes
\bea
\langle W_{\psi_1}  \rangle\Big|_{\rm 2loop} && =  \langle W_{\psi_2}  \rangle\Big|_{\rm 2loop} 
 \\ && = 1 + \frac{1}{24 k^2} \left( 2N^2  +1 \right)   \sim  \frac{1}{12} \left( \frac{N}{k} \right)^2 \non
\eea
\beq
 \langle W^{(1,2)} \rangle|_{\rm 2loop} = 1 +  \frac{1}{24 k^2} \left( 5N^2  +1 \right)   \sim  \frac{5}{24} \left( \frac{N}{k} \right)^2 
\non
\eeq
 
 \section{Cohomological equivalence at quantum level}
 \label{section4}
 
 It is easy to generalize the results (\ref{2loopresult}) to a generic $A$ site ($ A=0, \cdots , 2r-1$) and write
\bea
\label{WLF}
&& \langle W_{\psi_i}^{(A)} \rangle|_{i=1,2} =  1 -\frac{1}{24 k^2} \Big[ N_{A}^2 + N_{A+1}^2 - N_{A} N_{A+1} -1 \non \\ 
&& \qquad \qquad \qquad \,\, - 3 \frac{N_{A-1} N_{A}^2 + N_{A+2} N_{A+1}^2}{N_{A}+N_{A+1}} \Big]  + \cdots 
\eea
Similarly, generalizing result (\ref{bosonicresult}), for bosonic WL related to the $A$ site we have
\bea
&& \langle W^{(A)} \rangle = 1 -
\\
&& \qquad\frac{1}{24 k^2} \left[ N_{A}^2 - 3 N_{A-1} N_{A}- 3 N_{A} N_{A+1} -1 \right] + \cdots
\non
\label{WLB2}
\eea
 Exploiting these results it is interesting to understand how the classical cohomological equivalence (\ref{coho}) gets enhanced at quantum level. In fact, comparing the previous expressions one can easily realize that the following identity holds 
\bea
\label{identity}
&& \langle W_{\psi_i}^{(A)} \rangle_0 = e^{-i\frac{\ell_A}{2k} (N_{A} - N_{A+1})} \times  
\\
&&    \frac{N_{A} \, e^{i\frac{\ell_A}{2k} N_{A}} \langle W^{(A)} \rangle_0 +
N_{A+1} \, e^{-i\frac{\ell_A}{2k} N_{A+1}} \langle W^{(A+1)} \rangle_0}{N_{A} + N_{A+1}} \non
\eea
where $\ell_A = (-1)^{A+1}$ and the subscript ``0'' means perturbative result (framing zero).  Therefore, if we define ``framing--one'' quantities 
\bea
\label{framing}
&& \langle W_{\psi_i}^{(A)} \rangle_1 = e^{i \frac{\ell_A}{2k} (N_{A} - N_{A+1})} \, \langle W_{\psi_i}^{(A)} \rangle_0 \qquad j=1,2
\non \\
&& \langle W^{(A)} \rangle_1 =e^{-i\frac{\ell_A}{2k} N_{A}} \langle W^{(A)} \rangle_0
\eea
the previous identity can be rewritten as 
\bea
\label{coho2}
\langle W_{\psi_i}^{(A)} \rangle_1 =\frac{N_{A}  \langle W^{(A)} \rangle_1 + N_{A+1}\langle  W^{(A+1)} \rangle_1}{N_{A} + N_{A+1}}
\eea
and looks exactly like the classical relation (\ref{coho}).

\section{Matrix Model result at weak coupling} \label{section5}
\noindent
We now discuss the matrix model   for  the  necklace quiver theory described  in Section \ref{section2}.
The putative matrix integral, which  yields the partition function, can be easily obtained by combining  the basic building blocks given in \cite{Kapustin:2009kz}. We find \cite{Marino:2012az}
\be
\label{matrixmodel}
\mathcal{Z}\!=\! \mathcal{N}\!\!\int\!\! \prod_{B, i} {d\lambda_{B i} } e^{2 i k \ell_B \lambda_{Bi}^2}\!
\prod_{B=0}^{2r -1} \frac{\prod_{i<j} \sinh^2\left(\lambda_{Bi}-\lambda_{Bj}\right)}{\prod_{i,j}\cosh\left(\lambda_{Bi}-
\lambda_{B+1,j}\right)}
\ee
The constant $\mathcal{N}$ is an overall normalization, whose explicit form is irrelevant in our computation.

In the matrix model  language the $1/2$ BPS Wilson loop is not a fundamental object, as it can be computed  from 
the $1/4$ BPS Wilson loop through the cohomological relation \eqref{coho2}.
Therefore we focus on the latter.  It is given by the vacuum expectation value of  the following matrix
observable
\begin{align}
\label{Wb}
\!\!
W^{(A)}\!=&\frac{1}{N_A}\sum_{i=1}^{N_A} e^{ 2  \lambda_{Ai}}=1\!+\!\frac{2}{N_A} \text{Tr}(\Lambda_A)\!+\!\frac{2}{N_A } \text{Tr}(\Lambda^2_A)+\!\nonumber\\ &+\frac{4}{3 N_A } \text{Tr}(\Lambda_A^3)+\frac{2}{3N_A} \text{Tr}(\Lambda^4_A)+
O\left(\Lambda_A^5 \right)
\end{align}
where we have introduced the diagonal matrix  $\Lambda_A\equiv\text{diag}(\lambda_{A1},\cdots,\lambda_{A N_A})$ for future convenience.  In  the r.h.s. of \eqref{Wb} we can actually neglect  all the odd powers in $\Lambda_A$  since their expectation value vanishes at all order in $\frac{1}{k}$  due to  the symmetry property of the integrand  in  \eqref{matrixmodel} under  the parity transformation  $\lambda_{Ai}\to -\lambda_{Ai}$.

In order to construct the perturbative series for  $W^{(A)}$, first we rescale the eigenvalues $\lambda_{Ai}$  with  $\frac{1}{\sqrt{k}}$. Therefore, the measure factor for large $k$ reads
\begin{align}
\label{measure}
&\prod_{B=0}^{2r -1} \frac{\prod_{i<j} \sinh^2\frac{\lambda_{Bi}-\lambda_{Bj}}{\sqrt{k}}}{\prod_{i,j}\cosh\frac{\lambda_{Bi}-
\lambda_{B+1,j}}{\sqrt{k}}}=
\\
& \left[1+\frac{1}{k} \sum_{B=0}^{2r-1} P_B+O\left(\frac{1}{k^2}\right) \right]\prod_{B=0}^{2r -1} \prod_{i<j} \frac{(\lambda_{Bi}-\lambda_{Bj})^2}{k}
\non
\end{align}
where 
\begin{align}
P_B\equiv &\frac{1}{3}( N_B \mathrm{Tr}(\Lambda_B^2)-\mathrm{Tr}(\Lambda_B)^2)-
\frac{1}{2 }( N_{B+1}\! \mathrm{Tr}(\Lambda_B^2)\!+\nonumber\\
&+\!N_{B}\! \mathrm{Tr}(\Lambda_{B+1}^2)\!-\!2\mathrm{Tr}(\Lambda_B)\mathrm{Tr}(\Lambda_{B+1})).
 \end{align}
Since we shall write the final result as a  combination of vacuum expectation values in the Gaussian matrix model, we  have  chosen to use  the usual  Vandermonde determinant as the reference measure.  Moreover 
we have not  explicitly written  $\tfrac{1}{k^2}$ terms since they do not affect the final result. In fact, they cancel out with the normalization provided by the partition function. 

With the help of the expansion \eqref{measure}, it is straightforward to write down the expectation value of the  Wilson loop $W^{(A)}$ in terms of $P_B$ and $\Lambda_A$. We find
\be
\label{WWW}
\begin{split}
&\langle W^{(A)}\rangle=
1\!+\!\frac{2}{N_A k} \langle \mathrm{Tr}(\Lambda^2_A)\rangle
\!+\! \frac{1}{N_A k^2}\left[  \frac{2}{3} \langle \mathrm{Tr}(\Lambda^4_A)\rangle+\right.\\ &\left.+2\sum_{B=0}^{2r-1}\left[\langle  \mathrm{Tr}(\Lambda^2_A) P_B\rangle-\langle  \mathrm{Tr}(\Lambda^2_A) \rangle\langle P_B\rangle\right]\right]\!+\!O\left(\frac{1}{k^3}\right),\\
 \end{split}
\ee
where all the expectation values in the r.h.s. of  eq. \eqref{WWW} are taken in Gaussian matrix model of coupling constant $(-2 i  \ell_B)$. At the order $\frac{1}{k^2}$ the effect  of the interactions is entirely encoded in the   combination $\langle  \mathrm{Tr}(\Lambda^2_A) P_B\rangle-\langle  \mathrm{Tr}(\Lambda^2_A) \rangle \langle P_B\rangle$.  However this  combination vanishes unless $B=A-1$ or $B=A$ and thus the Wilson loop receives contributions from the nodes $A-1$, $A$ and $A+1$ (we recall that $P_B$ also depends on $\Lambda_{B+1}$). This 
is similar to what occurs in ABJ theories with the difference that  the node $A-1$ and $A+1$ are identified  there.

Using known results on  the expectation values of  $\mathrm{Tr}(\Lambda^n)$ and on correlators of traces in the Gaussian matrix model, we finally find
\begin{align}
&\langle W^{(A)}\rangle=
1+\frac{i \ell_A N_A}{2 k}-\\
&-\frac{1}{24 k^2}(4 N_A^2-3 N_{A-1}
   N_A-3 N_{A+1} N_A-1)+O\left(\frac{1}{k^3}\right)\nonumber
\end{align}
This expression coincides with the perturbative  result  for ${1/4}$ BPS Wilson loop given in \eqref{bosonicresult} dressed with the phase (\ref{framing}) corresponding to framing $1$.  With the help of cohomological relation \eqref{coho2}, we can also build $\langle W^{(A)}_{\psi_i} \rangle$ and  we find again the same result \eqref{WLF} of the perturbative computation.

\section{Discussion and perspectives}
\label{section6}

We have studied the two-loop perturbative behavior of the 1/2 BPS Wilson loop operators $W_{\psi_1}$ and $W_{\psi_2}$  introduced in \cite{Ouyang:2015qma, Cooke:2015ila} in the case of  $\mathcal{N}=4$ Chern-Simons matter quiver theories with alternating levels. 

The Feynman diagram analysis of Section \ref{section3} has shown that up to two loops the expectation values of $W_{\psi_1 }$  and $W_{\psi_2}$ are coincident and match the prediction from the perturbative expansion of the matrix model obtained in Section \ref{section4}.  Remarkably, in the field theory computation the coincidence between the two Wilson loops is true not only for the full result but it holds also for each of the contributing diagrams. At this perturbative order the two Wilson loops share exactly the same properties. We thus have to go up to three loops to look for hints of a possible lifting of the degeneracy between the two classically equivalent 1/2 BPS operators.

\begin{figure}
\centering
 \includegraphics[width=0.11\textwidth]{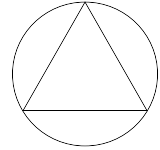}
\caption{Example of a three-loop diagram yelding different results for $W_{\psi_1 }$  and $W_{\psi_2}$.}
    \label{3loop}
\end{figure}

Indeed, at three loops some distinctive features in the perturbative computation arise.  First of all,  at this order  the two Wilson loops start giving different results at the level of single  diagrams. It is easy to find examples of this behaviour and  we provide the simplest one in Fig \ref{3loop}.
 
Evaluating the diagram for the two Wilson loops we obtain
\beq
W_{\psi_2 }|_{\rm Fig. 2}= - W_{\psi_1 }|_{\rm Fig. 2}  
\eeq
with
\begin{align} \label{3loopsc}
& W_{\psi_1 }|_{\rm Fig. 2}   = - \frac{2}{k^3} \frac{ N_1 N_2 (N_1^2 + N_2^2+2)}{N_1+N_2}  \,\, \textrm{I}(\epsilon)  \\ 
& \textrm{I}(\epsilon)  =\frac{4}{3}\,\Gamma\left(-\tfrac{1}{2}+3\epsilon\right)
\left[\frac{\Gamma(\tfrac{1}{2}-\epsilon)\Gamma(1+\epsilon)}{(4\pi)^{1-\epsilon}\Gamma(1+2\epsilon)}\right]^3\underset{\e \rightarrow 0}{\longrightarrow}
 -\frac{1}{24\pi}  
\non
\end{align}
The extra minus sign in $W_{\psi_2}$ compared to $W_{\psi_1}$ comes from the different scalar couplings in the two Wilson loop definitions. 
This situation is very similar to what happens for the one-loop fermion exchange contribution of Section \ref{section3}. However, while in that case the diagram is eventually  discarded because the integral has been shown to be $\mathcal{O}(\epsilon)$, in the present case a finite contribution survives in the $\e \to 0$ limit.

Another source of possible differences might come from the Yukawa vertices in the potential, which start contributing at three loops. In fact, while minimal couplings entering up to two loops are diagonal in the flavour space, Yukawa vertices are in general flavour changing and the computation might become sensible to the flavour choice of the spinor insertions on the contour.  

Based on these general observations, we expect a different result for a subset of three--loop Feynman diagrams and it would be crucial to check if the differences are compensated when we sum over all the contributions. If this were the case, the common result of the two operators should match the three-loop expansion of the matrix model. Instead, if the differences would not cancel against each others, 
and cannot be absorbed in a change of framing, the prediction from the matrix model could be matched only by a specific linear combination of the two Wilson loops, as suggested in \cite{Cooke:2015ila}. A more radical possibility is that  no linear combination satisfies the constraint and therefore the cohomological equivalence is broken at the quantum level.
We will report on the ongoing three--loop analysis in \cite{forthcoming}.  

Moreover, it would be interesting to understand how the $W_{\psi_i }$ operators of the  $\mathcal{N}=4$ models fit in the   
family of  1/2 BPS Wilson loops  recently introduced \cite{Ouyang:2015iza} for general $\mathcal{N}=2$ theories, where a perturbative analysis such as the one completed in this paper could also be applied.


\section{Acknowledgements}

This work has been supported in part by MIUR, INFN and MPNS-COST Action MP1210 ``The String Theory Universe".

\end{document}